\RequirePackage[2020-02-02]{latexrelease}
\documentclass[aps,superscriptaddress,twocolumn]{revtex4-1}
\usepackage{graphicx}
\begin{document}

\title{Controllable single spin evolution at sub-harmonics of electric dipole spin resonance enhanced by four-level Landau-Zener-St{\"u}ckelberg-Majorana interference}

\author{D.V. Khomitsky}
\email{khomitsky@phys.unn.ru}
\affiliation{Department of Physics, National Research Lobachevsky State University of Nizhny Novgorod, 603950 Gagarin Avenue 23, Nizhny Novgorod, Russian Federation}
\author{M.V. Bastrakova}
\affiliation{Department of Physics, National Research Lobachevsky State University of Nizhny Novgorod, 603950 Gagarin Avenue 23, Nizhny Novgorod, Russian Federation}
\affiliation{Russian Quantum Center, 143025 Skolkovo, Moscow,  Russian Federation}
\author{V.O. Munyaev}
\affiliation{Institute of Information Technologies, Mathematics and Mechanics, National Research Lobachevsky State University of Nizhny Novgorod, 603950 Gagarin Avenue 23, Nizhny Novgorod, Russian Federation}
\author{N.A. Zaprudnov}
\affiliation{Department of Physics, National Research Lobachevsky State University of Nizhny Novgorod, 603950 Gagarin Avenue 23, Nizhny Novgorod, Russian Federation}
\author{S.A. Studenikin}
\email{sergei.studenikin@nrc-cnrc.gc.ca} 
\affiliation{Emerging Technologies Division, National Research Council of Canada, Ottawa, ON K1A0R6, Canada}

\begin{abstract}
Sub-harmonics of electric dipole spin resonance (EDSR) mediated by Landau-Zener-St{\"u}ckelberg-Majorana (LZSM) tunneling transitions are studied numerically and analytically in a Zeeman-split four level system with strong spin-orbit coupling that can be realized, for example, in a GaAs-based double quantum dot in a single-hole regime. The spin qubit is formed in one of the dots and the second dot is used as an auxiliary element to enhance functionality of the spin qubit. In particular, it is
found that the spin rotation rate can be essentially enhanced due to the tunnel coupling with the auxiliary dot on both the main EDSR frequency and at its high sub-harmonics allowing the coherent spin $\pi$-rotations on a 10-ns time scale.
Spin manipulation on high sub-harmonics is promising for new time-efficient schemes of the spin control and readout in qubit devices operating at high magnetic fields where the main harmonic is inaccessible due to hardware limitations.
\end{abstract}

\date{\today}
\maketitle

\section{Introduction}

It is known that the effects stemming from the Landau-Zener-St{\"u}ckelberg-Majorana (LZSM) interference during the close passage of energy levels \cite{Nori2010,Nori2023,DiGiacomo2005,Nori2018,Ludwig2014,Ludwig2015} and the associated multi-photon effects \cite{Delone,GrifoniHanggi1998} can be  revealed in a number of systems including the  single or double semiconductor quantum dots (QD). Here the charge \cite{StehlikNori2012,Ludwig2013,Raikh2022,Burdov2004,Burdov2005}, spin \cite{Rashba2003,GolovachLoss2006,Koppens2006,Nowack2007,NadjPerge2010,NadjPerge2012,Ludwig2015b,Petersson2012,StehlikPetta2014,Nowak2012,Manchon2022,StehlikPetta2016,Benito2019,Nori2019,Studenikin2019,Studenikin2022,Studenikin2023,Loss2023,Lasek2023,Platero2023}, and valley \cite{Szafran2017,Petta2018,Burkard2022,Klemt2023} degrees of freedom can be involved into the LZSM interference. The undoubtful advantage of the LZSM interferometry is a flexible setup of the system state by tuning the external field parameters allowing at the same time the LZSM spectroscopy of both charge and spin dynamics. Comparing to the well-understood two-level case \cite{Nori2010,Nori2023,DiGiacomo2005} the multi-level systems demonstrate much more complicated behavior under periodic driving both for solid state structures  \cite{Rudner2008,Schreiber2011,Oosterkamp1998,Khomitsky2012,Budagosky2016,Granger2015,Studenikin2018,Studenikin2021,Gomez2019,Grimaudo2020,Kitamura2020,Chen2021,Malla2021,Zhou2014,Pasek2018}, interacting Josephson qubits \cite{SataninNori2012,SataninNori2014,Bastrakova2021,He2023}, atomic systems \cite{DiGiacomo2005,Vasilev2007,Novelli2015,Liang2020,Liu2021,Nath2020,Nath2021} and systems where the effects of the dissipation are significant \cite{Nori2023,Zhang2023,Bonifacio2020}.
The effective coupling between charge and spin states induced by the spin-orbit interaction (SOI) has been manifested in the electron systems \cite{Sherman2012,Sherman2018} including the three-level model of a hybrid qubit \cite{Zhou2023b}, in hole spin devices \cite{Studenikin2018,Studenikin2019,Studenikin2022,Studenikin2023,Loss2023} and in narrow band-gap semiconductors \cite{NadjPerge2010,NadjPerge2012,Petersson2012,StehlikPetta2014,StehlikPetta2016} including the observation of a variety of complex LZSM patterns. In particular, the generation of both integer higher or fractional (sub) harmonics in quantum dots including their manifestation in the Electric Dipole Spin Resonance (EDSR) has been reported \cite{StehlikPetta2014,Nowak2012,Szafran2017,Manchon2022,Vandersypen2015,Burkard2015,Zhou2023}. It is known that the sequence of frequencies 
\begin{equation}
\omega=n \omega_0,
\label{harm}
\end{equation}
 where $n$ is an integer and $\omega_0$ is the primary frequency is called harmonics of $\omega_0$ and the sequence of fractional multiples 
 $\omega=\omega_0/k$ satisfying
\begin{equation}
k \omega=\omega_0,
\label{subharm}
\end{equation}
where $k$ is an integer is called sub-harmonics or undertones of $\omega_0$ \cite{Partch1974}. 

In our present work we will focus on spin dynamics at sub-harmonics of the EDSR at chosen driving frequency $\omega$ satisfying (\ref{subharm}) where the fundamental frequency $\omega_0$ equals the Zeeman level splitting $\Delta_Z$ and is scaled linearly with the applied magnetic field. The condition (\ref{subharm}) here reads as $k \omega=\Delta_Z$ where $k$ is an integer (we choose $\hbar =1$ here and below). The high sub-harmonic numbers $k$ in (\ref{subharm}) will correspond to a growing magnetic field at fixed driving frequency $\omega$. The condition (\ref{subharm}) can be also interpreted as the multi-photon process when the integer quanta $k$ of photons with frequency $\omega$ matches the desired splitting $\omega_0$ \cite{Delone,GrifoniHanggi1998}. In our system, however, the LZSM effects stemming from the dynamical level intersection during the periodic driving modify the classical $k$-photon power law for transition probability making it strongly interference-dependent and comparably high on many sub-harmonics.

Originally, the spin rotations due to a periodical spatial displacement in the presence of SOI were reported in Refs. \cite{Rashba2003,GolovachLoss2006,Koppens2006,Nowack2007,NadjPerge2010,NadjPerge2012,Ludwig2015b}.
Here we continue our research \cite{KS2022,KhomitskySemic22} of the spin state evolution  both within the Floquet stroboscopic technique and in the continuous time. In this paper we focus on the enhanced spin-flip rotations at high EDSR sub-harmonics stemming from the interplay between spin rotations in one dot and the spin flip LZSM-like  transitions to spin states formed in the other dot due to the strong SOI. 
The goal of this work is to investigate new regimes of hole electric dipole spin resonance (EDSR) enhanced by the LZSM tunneling transitions, in particular, to theoretically describe both analytically and numerically the electric field-driven spin evolution in new, specially chosen  parameters of the quantum system at the sub-harmonic frequency regimes that can be compared to the standard EDSR methods   \cite{Koppens2006,Nowack2007,NadjPerge2010,NadjPerge2012,Ludwig2015b,Sherman2012,Sherman2018,Zhou2023b}. The main achievement of the present work is the analytical and numerical demonstrations of the high-sub-harmonics (\ref{subharm}) EDSR that may be useful for achieving high Zeeman energy regimes, for example, in high magnetic fields that would require the terahertz frequency radiation.  

 The parameters of our modeling are close to experimental conditions \cite{Studenikin2018,Studenikin2019} which include a relatively weak tunnel coupling of the order of $1\ldots 4$ $\mu eV$. We consider the spin-dependent tunneling and single-spin rotations in a periodically driven system of four spin levels with SOI. The EDSR sub-harmonics for $k>1$ may have an important experimental and technological applications since they provide an ability to obtain the desired spin manipulations at higher magnetic fields than for $k=1$, or, equivalently, for higher effective frequencies. For example, manipulating the spin on the driving frequency of 10 GHz at sub-harmonic $k=10$ is equivalent to the spin manipulation at the basic sub-harmonic $k=1$ but for much higher frequency of 100 GHz, which may be not accessible due to hardware limitations. Our modeling shows that the spin dynamics at the EDSR sub-harmonics is very sensitive to tunneling for the case of strong SOI even for weak tunneling coupling between the dots. We also demonstrate the ability of controllable rotations of the spin state on the Bloch sphere around x- and y- axis.  
  
This paper is organized as follows. In Sec. II we briefly summarise the principal properties of the Hamiltonian and the observables described in more detail in Ref.\cite{KS2022}. In Sec. III we discuss the primary regimes of tunneling and spin flip in terms of the associated resonance conditions and focus on the observation conditions for the EDSR sub-harmonics. For the overall validity  check we also present a simple analytical approximation of spin evolution within the RWA approximation applied to a simplified four-level scheme. In Sec. IV we describe the modeling with the numerical parameters and describe the EDSR sub-harmonics on the maps of averaged spin-dependent tunneling probability, comparing them with the analytical results obtained in the RWA. In Sec. V we calculate the maximally achievable spin flip amplitudes and spin flip rates on EDSR sub-harmonics and present the spin evolution examples for the regimes of ``pure'' and ``hybrid'' EDSR. Finally, in Sec. VI we summarize the results.

\section{Four-level model for two Zeeman-split states}

The model we use has been largely derived in our preceding paper \cite{KS2022} and is based on the solution of the non-stationary Schr{\"o}dinger equation with the Hamiltonian for the 1D double quantum dot with SOI and subject to constant magnetic and periodic electric fields:
\begin{equation}
H=H_{\rm{2QD}}+H_Z+H_{\rm{SO}}+V(x,t).
\label{ham}
\end{equation}

The Hamiltonian (\ref{ham}) is written for quantum states in 1D space containing two minima of neighboring QDs and the tunneling barrier between them which corresponds to the layout of the experiments \cite{Studenikin2018} were such structure  was formed in a two-dimensional hole gas by the electrostatic gates and the tunneling took place in one dimension across the barrier between the dots. The Hamiltonian (\ref{ham}) is written in a single-particle approximation, although the models with two electrons or holes have also been developed \cite{Taylor2007,Platero2022}. 
In Eq.~(\ref{ham}) $H_{\rm{2QD}}=k_x^2/2m+U_0(x)$ with $U_0(x)=U_0((x/d)^4-2(x/d)^2)$ is the hole Hamiltonian with the effective mass $m$ in the lowest subband of size quantization (hereafter we use units with $\hbar=1$). The double well potential $U_0(x)$ with height $U_0$ is characterized by the interdot center distance $2d$. The Hamiltonian includes the Zeeman splitting term 
\begin{equation}
H_Z=\frac{1}{2}g\mu_B B_z \sigma_z 
\label{hz}
\end{equation}
generated by the constant magnetic field along which the Oz axis is chosen and g is the effective hole g-factor. This term determines the Zeeman splitting 
\begin{equation}
\Delta_Z=g\mu_B B_z. 
\label{deltaz}
\end{equation}

The term $H_{\rm{SO}}$ in Eq.(\ref{ham}) corresponds to the contribution from bulk Dresselhaus SOI linear in the wavevector which is the leading term for GaAs-based low-dimensional structures \cite{GolovachLoss2006} 
\begin{equation}
H_{\rm{SO}}=\beta_D \sigma_x k_x, 
\label{hso}
\end{equation}
here $\beta_D$ is the strength of SOI. The non-stationary term $V(x,t)$ in Eq.~(\ref{ham}) contains both the static detuning and the periodic driving by the electric field. According to the experiments \cite{Studenikin2018} for $t<0$ only the static detuning is present,
\begin{equation}
V(x,t<0)=U_d f_d(x).
\label{vxtl0}
\end{equation}

\begin{figure}[tbp]
\centering
\includegraphics[width=0.45\textwidth]{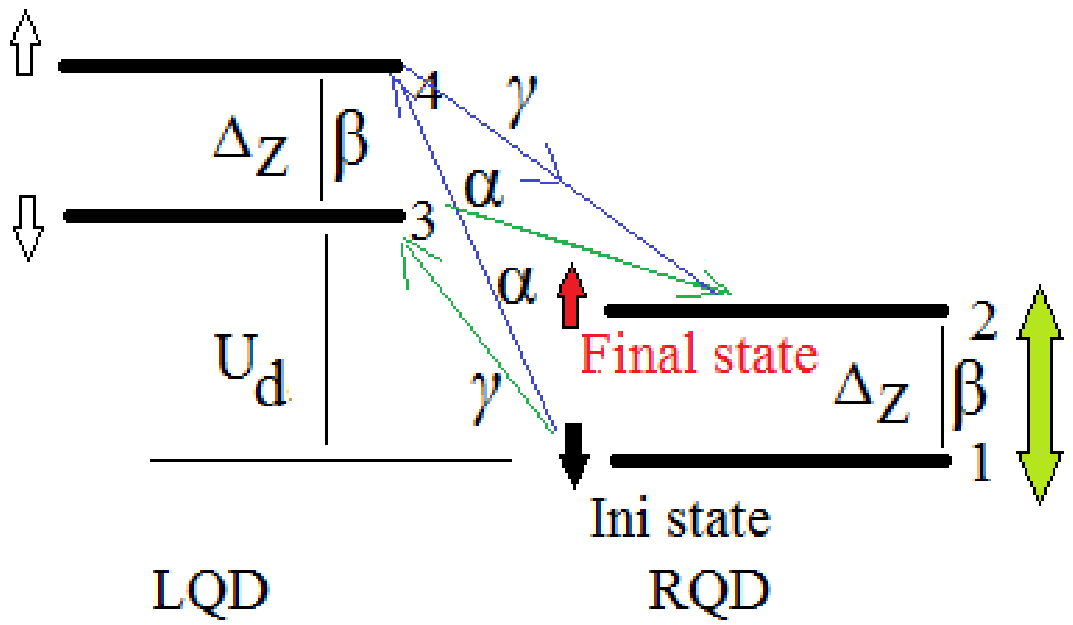}
\caption{Principal subspace of a double QD: a four-level system with the Zeeman doublet splitted by $\Delta_Z$ in left and right QDs which bottoms are shifted by the detuning $U_d$. The initial state (black arrow) is the spin-down state in the right QD and the final state in the right QD is marked by the red arrow. The principal transitions triggered by the driving field (double vertical arrow) are labeled by angled arrows and include (a) spin-conserving tunneling during the LZSM level passage with rate $\gamma$; (b) spin-flip tunneling during the same passage with rate $\alpha$; (c) EDSR in a single QD without tunneling with the rate $\beta$.} 
\label{figlevels}
\end{figure}

In (\ref{vxtl0}) the function $f_d=(x/d_1)^3-3/2\cdot (x/d_1)^2$ models the smooth connection between the detuning and the double well potential with $d_1=1.5 d$. This function produces a bottom-down shift linear in $U_d$ for the right QD at $U_d<0$ and the the bottom-up shift at $U_d>0$, giving a convenient model of detuning. The sum of the double well potential $H_{\rm{2QD}}$ and the detuning (\ref{vxtl0}) creates the profile with the lowest subspace of two pairs of Zeeman-split levels  in each of the two quantum dots as it is shown in Fig.\ref{figlevels}. 
We diagonalize the time-independent part of the Hamiltonian (\ref{ham}) and obtain the energy levels $E_n$ and the eigenfunctions $\phi_n(x)$, the latter being two-component spinors. 

From the moment of time $t=0$ the periodic driving is turned on corresponding to the time-dependent potential
\begin{equation}
V(x,t \ge 0)=\left[U_d+V_d \sin \omega t \right]f_d(x).
\label{vxtg0}
\end{equation}

The time-dependent Schr{\"o}dinger equation is solved in a standard fashion with solution expressed as a sum of the eigenfunctions of the stationary part with time-dependent coefficients:
\begin{equation}
\psi(x,t)=\sum_n C_n(t) e^{-i E_n t} \phi_n(x).
\label{psixt}
\end{equation}

In the present work we follow the approximation from Ref.\cite{KS2022} and perform our modeling in the subspace of four lowest levels $E_1, E_2, E_3, E_4$ shown in Fig.\ref{figlevels} since higher levels have less than 10 percent contribution. 
The small observed impact of higher levels is due to their distant location (about 2 meV and higher) from the ground manifold which makes them weekly involved into the dynamics for the driving amplitude $V_d=30 \dots 75$ $\mu eV$ considered in our modeling. 
So, the primary regimes of the evolution observed in the experiments \cite{Studenikin2018,Studenikin2021} can be described in a four-level approximation. In each QD a ground pair of spin-resolved levels $E_1$, $E_2$ and $E_3$, $E_4$ is taken, with opposite z-projections of spin in the presence of the Zeeman term ({\ref{hz}}) which is marked by the vertical arrows near the levels in Fig.\ref{figlevels}. 
This manifold of four levels represents the minimal model of the double dot system with the charge and spin degrees of freedom. The system of the differential equations for the coefficients $C_n(t)$ is accompanied by the initial condition $C_n(0)$ describing the spin-down wavepacket injected into the ground level of the right QD which resembles the experimental conditions \cite{Studenikin2018}. We solve the equations for $C_n(t)$ both within the continuous time and in the Floquet stroboscopic representation. In the latter case the results are shown at the time moments $t=NT$ where $T=2\pi / \omega$ is the driving field period. Our observables are calculated using the reconstructed wavefunction (\ref{psixt}) across the whole double QD at the stroboscopic time moments, $t=NT$. We begin with the time-dependent probability $P_{L}(t)$ to find the particle in the left QD which corresponds to the tunneling efficiency,
\begin{equation}
P_{L}(t)=\int_{-\infty}^{0} \mid \psi(x,t) \mid^2 dx.
\label{pl}
\end{equation}

According to the experiments \cite{Studenikin2018} the electrical current through the double QD is proportional to the $P_{L}$  averaged over the whole observation time so we calculate the time average of (\ref{pl}):
\begin{equation}
P_{L}=\frac{1}{NT} \int_0^{NT} P_{L}(t) dt.
\label{psaver}
\end{equation}

The next observable is the spin z-projection which enters as another degree of freedom measured in left or right QD, respectively:
\begin{equation}
\sigma_z^{L}(t)=\int_{-\infty}^{0} \langle \psi |\sigma_z| \psi \rangle dx,
\label{sigmazl}
\end{equation}

\begin{equation}
\sigma_z^{R}(t)=\int_{0}^{\infty} \langle \psi |\sigma_z| \psi \rangle dx.
\label{sigmazr}
\end{equation}

Because of the presence of SOI term (\ref{hso}) the evolution of the spin demonstrates that it may be no longer conserved and the contributions (\ref{sigmazl}), (\ref{sigmazr}) may be not coupled via the normalization condition. In the present paper we focus mainly on the spin dynamics in a particular (right) dot working as a spin qubit and the left dot is used as an additional auxiliary element to improve performance of the right-dot spin qubit and realize new functionalities, the sub-harmonics EDSR operations. In our simulations the right dot spin qubit is initialized in the spin-down state, that can be realized experimentally as well \cite{Studenikin2023}.
As in (\ref{psaver}) we calculate the time-averaged spin z-projection in left and right QD as following:
\begin{equation}
\sigma_z^{L,R}=\frac{1}{NT} \int_0^{NT} \sigma_z^{(L,R)}(t) dt.
\label{szaver}
\end{equation}

Our modeling shows that a typical steady map of averaged observables (\ref{psaver}), (\ref{szaver}) is formed on a scale of $N=100 \ldots 500$ periods of the driving field, $T$, which is the observation interval in our model.

\section{Regimes of dynamics and analytical approximation}

\subsection{Regimes of dynamics}

Various transitions can be triggered by the periodic electric field with the potential $V(x,t)$ in a four-level system as depicted in Fig.\ref{figlevels} by arrows connecting different states both in a single QD and in the neighboring QDs. 
The resonance effects of periodic driving on tunneling can be described with participation of certain number of photon quanta $k \omega$, the phenomenon referred as the Photon Assisted Tunneling (PAT), 
or multi-photon absorption regime \cite{Nori2010,Nori2023,Delone,GrifoniHanggi1998}. 
The regime of evolution can be described by the adiabaticity parameter \cite{Nori2010,Nori2023,DiGiacomo2005}
\begin{equation}
\delta=\frac{\Delta^2}{4 v},
\label{adiabpar}
\end{equation}
where $\Delta$ is the level coupling being basically described by the tunneling strength and $v=d(E_2-E_1)/dt$ is the rate of distance change between the interacting levels in energy space which can be estimated for periodic driving as $v=\omega V_d$. For our model typical values are $\Delta \sim 1$ $\mu eV$, $\omega \sim 10$ $\mu eV$, $V_d \sim 100$ $\mu eV$ so the adiabaticity parameter $\delta \sim 10^{-3} ... 10^{-4}$ which corresponds to the fast passage of levels. In such case the two-level occupation probability patterns in the parameter space can be well described by analytical approximations \cite{Nori2010,Nori2023} that match well the numerical results for our model \cite{KS2022}. In the present paper we will focus on the EDSR sub-harmonics affected by tunneling which require the approximations based on a four-level model where only limited analytical approximations are available which are considered in the next Subsection. 

The following regimes of driven evolution can be identified in Fig.\ref{figlevels} where the initial and final states assigned to the spin-up or spin-down level in the corresponding QD are labeled by the black and red arrow, respectively, indicating the spin projection.
The arrows connecting the states (1, 3) or (4, 2) in Fig.~\ref{figlevels} depict the spin-conserving tunneling with the rate $\gamma$ when a number of photon quanta equals the detuning amplitude $|U_d|$ that corresponds to the PAT:
\begin{equation}
|U_d|=k_1 \omega.
\label{pat} 
\end{equation}

Another regime is the spin-flip tunneling shown by the arrows connecting the states (1, 4) or (3, 2) in Fig.\ref{figlevels}. Here a number of photon quanta equals the detuning amplitude plus or minus the Zeeman splitting providing the hole to tunnel to the level with another spin projection with the rate $\alpha$: 
\begin{equation}
|U_d| \pm \Delta_Z=k_2 \omega.
\label{patflip} 
\end{equation}

In our modeling we consider the spin-down initial state in the right QD being the ground state 1 in Fig.\ref{figlevels} for the negative detuning which closely resembles the situation in recent experiments \cite{Studenikin2018}. For such initial spin state the spin flip tunneling connecting the states (1,4) in Fig.\ref{figlevels} and described by the plus sign in (\ref{patflip}) will dominate in the dynamics.

There is also a ``pure'' or single-dot EDSR without efficient tunneling between the states (1, 2) or (3, 4) in Fig.\ref{figlevels}. It is described by the rate $\beta$ and takes place when the EDSR condition is satisfied:
\begin{equation}
\Delta_Z=k_3 \omega.
\label{edsrcond}
\end{equation}

The value $k_3=1$ in Eq.~(\ref{edsrcond}) corresponds to  the main EDSR frequency when the driving frequency itself matches the Zeeman splitting (\ref{deltaz}) calculated in the presence of SOI. The values ($k_3=2,3,...$) in Eq.~(\ref{edsrcond}) describe the sub-harmonics of the EDSR which will be in the focus of the present paper. The maximum number $k_{3m}$ of the EDSR sub-harmonics which can be observed for a given driving frequency is obviously limited by the condition
\begin{equation}
k_{3m} \le \Delta_Z / \omega,
\label{maxk3}
\end{equation}
although the higher sub-harmonics satisfying (\ref{maxk3}) may have low intensity which will be discussed below. 
Another limitation of the maximal sub-harmonic number which can be derived from the level structure in Fig.\ref{figlevels} is associated with driving amplitude strength,
\begin{equation}
|U_d|+\Delta_Z \le V_d,
\label{maxdeltaz}
\end{equation}
which means that one needs strong enough driving $V_d$ to make the levels with opposite spin in neighboring dots feel each other during the periodic driving. This is a usual limiting factor in observing the LZSM patterns in (Driving, Detuning) plane \cite{Nori2010,Nori2023,Studenikin2018}.
Condition (\ref{maxdeltaz}) limits the Zeeman splitting from above and thus, together with (\ref{maxk3}), limits the number of EDSR sub-harmonics that can be observed at a fixed set of parameters.

A principal feature of the regimes with resonances (\ref{pat}) - (\ref{edsrcond}) is that they involve basically a pair of two levels in the dynamics with the specific rate $\gamma$, $\alpha$ or $\beta$ introduced above. As to the EDSR (\ref{edsrcond}), we call this regime  the ``pure'' EDSR when this only resonance condition (\ref{edsrcond}) is satisfied. Our previous modeling \cite{KS2022} has shown that the spin flip is still affected by the tunneling in this non-resonant case, although the tunneling itself is not effective. The results in Ref.\cite{KS2022} have shown that a combination of all three resonance conditions  (\ref{pat}) - (\ref{edsrcond}) is possible when the indices satisfy the relation $k_2=k_1+k_3$. The system in such regime is essentially a four-level one and cannot be described by simple two-level schemes. From the practical point of view it is important that the hybrid resonance points can always be found in the map of (Driving frequency, Magnetic field) parameters which will be considered in the following Section. The EDSR which takes place under such condition of the merging of the three resonances is characterized by significant involvement of tunneling between all four levels. We thus call it the ``hybrid'' EDSR and will compare the characteristics of its sub-harmonics with the ones for ``pure'' EDSR in the following Sections.

\subsection{Analytical approximation}

We perform an analytical estimates of the spin evolution regimes in our system using the minimal four-level model for the Zeeman split states 1 and 2 in the right QD and 3 and 4 in the left QD:
\begin{equation}
H =\left[ 
\begin{array}{cccc}
-\frac{\Delta_Z}{2}+V(t) & \beta & \gamma & \alpha \\ 
& \frac{\Delta_Z}{2}+V(t) & \alpha & \gamma \\
h.c. & & -\frac{\Delta_Z}{2} & \beta \\
& & & \frac{\Delta_Z}{2}
\end{array}
\right],  
\label{hsimple}
\end{equation}
where $V(t)=U_d+V_d \sin \omega t$ takes into account both detuning and driving applied to the states 1 and 2 in the right dot and h.c. labels the Hermitian conjugation. The Hamiltonian (\ref{hsimple}) is written in the subspace of the eigenstates of two isolated QDs with the lowest pair of spin-resolved levels in each of the dots at the absence of tunnel coupling, magnetic field, SOC and driving. The parameters $\gamma$ and $\alpha$ represent the coupling between the dots and define the spin-conserving and the spin-flip tunneling rates, respectively. The parameter $\beta$ determines the spin flip rate in the isolated dot. The nonstationary Schr\"odinger equation with Hamiltonian (\ref{hsimple}) in such representation is written for the vector $(C_1(t),C_2(t),C_3(t),C_4(t))$ giving the level occupations $|C_k(t)|^2$ for the corresponding state. The initial condition for the typical negative detuning when the ground state is the spin-down state in the right dot reads as $(1,0,0,0)$. The occupancy (\ref{pl}) of the left dot is given as $P_L(t)=|C_3(t)|^2+|C_4(t)|^2$ and the z-projection (\ref{sigmazr}) of the spin in the right dot
is given as $\sigma_z^R(t)=|C_2(t)|^2-|C_1(t)|^2$.

We are interested in the EDSR regime (\ref{edsrcond}) and its sub-harmonics so it is natural to introduce a small frequency detuning parameter
\begin{equation}
\delta_{\omega}=\Delta_Z-k_{3}\omega,
\label{deltarwa}
\end{equation}
which approaches zero on the EDSR and its sub-harmonics.
After performing the RWA approximation for the Schr\"odinger equation with Hamiltonian (\ref{hsimple}) \cite{Nori2010,Nori2023,SataninNori2012,SataninNori2014,Bastrakova2021}  in the limit $\delta_{\omega} \to 0$ and away of the other resonances (\ref{pat}) and (\ref{patflip}) we obtain the analytical expressions for the level occupancies $|C_k(t)|^2$ and for the z-projection of the spin. Its time average for the given initial condition $\psi(0)=(1,0,0,0)$ reads as following:
\begin{equation}
\langle \sigma_z^R \rangle = -1 + \sum_{k_3} \frac{\gamma_{k_3}^2}{\gamma_{k_3}^2+\left(\Delta_z^{(k_3)}-\Delta_Z\right) ^2}.
\label{szrwa}
\end{equation}

In (\ref{szrwa}) we introduce
\begin{equation}
\Delta_z^{(k_3)}=k_3 \omega-\frac{2\beta^2}{k_3 \omega}+\alpha^2 \sum_{k=-\infty}^{\infty}
\frac{J_{k+k_3}^2\left(\frac{V_d}{\omega}\right)-J_{k-k_3}^2\left(\frac{V_d}{\omega}\right)}
{U_d+k \omega}
\label{deltak3}
\end{equation}

and

\begin{equation}
\gamma_{k_3}=2\alpha \gamma \sum_{k=-\infty}^{\infty} J_k\left(\frac{V_d}{\omega}\right)
\frac{J_{k+k_3}\left(\frac{V_d}{\omega}\right)+J_{k-k_3}\left(\frac{V_d}{\omega}\right)}
{U_d+k \omega},
\label{gammak3}
\end{equation}
where $J_k(z)$ is the $k$-order Bessel function. 
From (\ref{szrwa}) - (\ref{gammak3}) one may expect that the intensity of higher sub-harmonics will decrease with increasing driving strength $V_d$ and/or decreasing driving frequency $\omega$. Indeed,  high numbers of sub-harmonics require the strong driving according to (\ref{maxk3}), (\ref{maxdeltaz}) for these subharmonics to be visible. Using the asymptotic representation of the Bessel functions $J_k (x) \sim \sqrt{2/\pi x} \cos(x-\pi k/2-\pi/4)$ with $x=V_d/\omega$ one can conclude that for growing $V_d$ or decreasing $\omega$ a slow power law decrease in sub-harmonic intensity following this asymptotic behavior is expected.
We will use the approximation (\ref{szrwa}) in the next Sec. for comparison with the full-scale computational simulation of the dynamics and the time averaged values of the observables.

\section{Numerical parameters and tunneling probability maps}

 We will obtain the numerical results for the hole GaAs DQD structure with the parameters similar to those in \cite{Studenikin2018,KS2022}: the hole effective mass $m_h=0.11 m_0$, the interdot minima distance $2d=116$ nm, the barrier height $U_0=4$ meV, the g-factor $g=1.35$ and the SOI Dresselhaus constant $\beta_D=3$ $\rm{meV} \cdot \rm{nm}$. Under these conditions and using the potential $V(x,t$ described is Sec. II  the calculated spin-conserving tunneling rate is about 2.2 $\mu eV$ and the spin-flip tunneling rate is about 1.0 $\mu eV$, that is close to the experimental conditions in Ref.\cite{Studenikin2018}. The initial state is the spin-down wavepacket with width $\sim d$ centered in the right QD represents the hole injected from the right lead to the ground state of the right QD in accordance with the experimental settings. In our study we employ two-dimensional (2D) maps of averaged spin-dependent tunneling probability (\ref{szaver}) in the plane of specifically chosen sets of parameters where the different regimes described in previous Sec. are identified and explored.

\subsection{Spin flip map and EDSR sub-harmonics in the plane of magnetic field and driving frequency}

\begin{figure}[tbp]
\centering
\includegraphics*[width=0.5\textwidth]{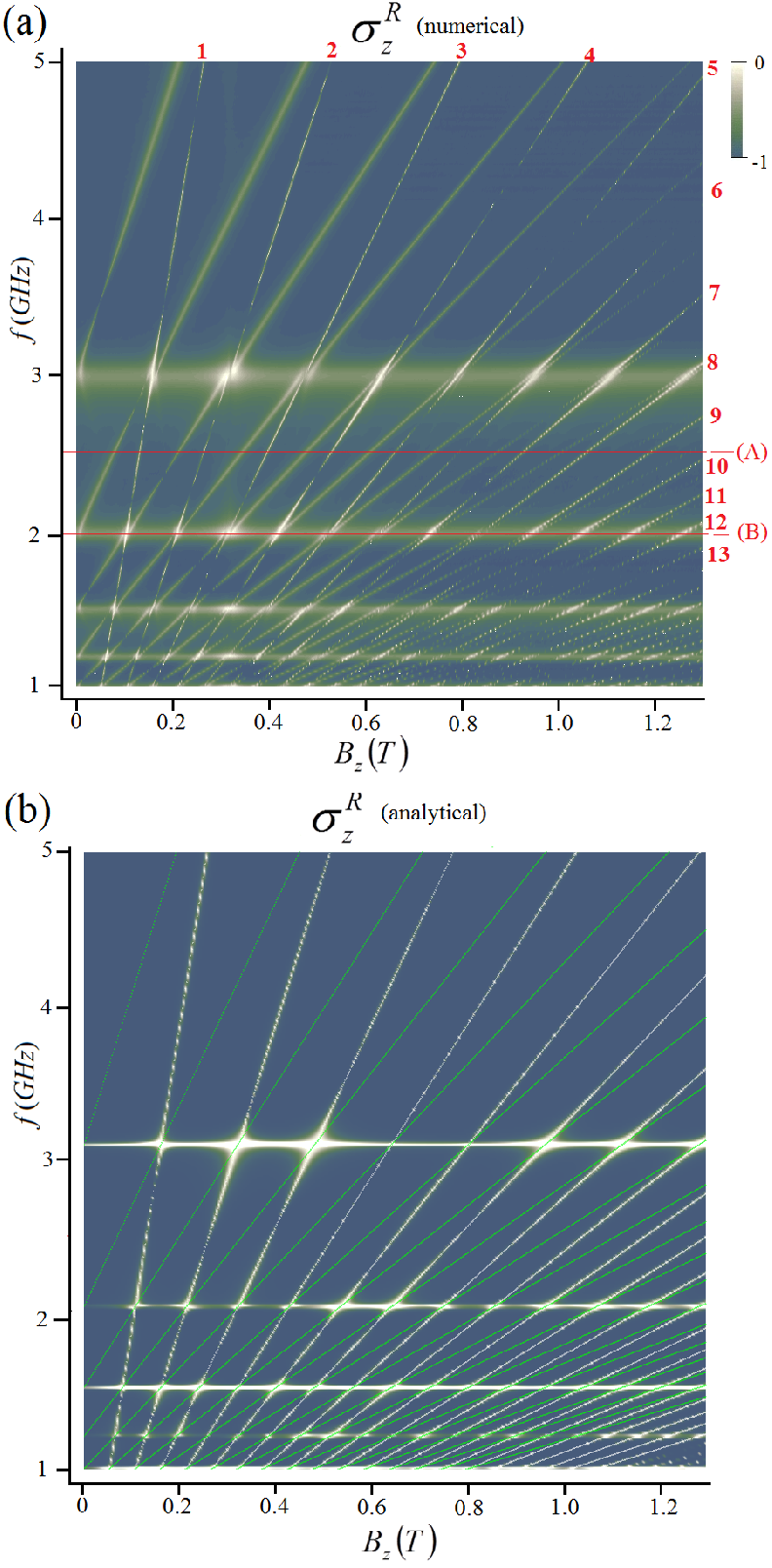}
\caption{Maps of $\sigma_z^R$ shown in $(B_z, f)$ plane at fixed detuning $U_d=-25$ $\mu eV$ and fixed driving amplitude $V_d=75$ $\mu eV$ calculated (a) numerically and (b) analytically within the RWA approximation (\ref{szrwa}). The steep angled lines on panel (a) labeled by red numbers are the EDSR sub-harmonics. Spin evolution on the EDSR sub-harmonics for the case of pure EDSR without effective tunneling (line A) at $f=2.5$ GHz and for the case of hybrid EDSR sub-harmonics where the spin flip is combined with strong tunneling (line B) at $f=2.05$ GHz will be considered in the next Sec. in time domain.}
\label{figbzf}
\end{figure}

A comprehensive picture of spin-dependent tunneling can be constructed by viewing onto the time-averaged observable (\ref{szaver}) in the $(B_z,f)$ plane. In Fig.~\ref{figbzf} we show the maps of $\sigma_z^R$ in $(B_z, f)$ plane at fixed detuning $U_d=-25$~$\mu eV$ and fixed driving amplitude $V_d=75$~$\mu eV$ calculated (a) numerically and (b) analytically within the RWA approximation (see Sec.IIIB). The spin-conserving tunneling (\ref{pat}) is shown by the horizontal lines and the spin-flip tunneling (\ref{patflip}) is shown by the first family of thick angled lines. The second family of the thin steepest angled lines labeled by red numbers represents the EDSR sub-harmonics (\ref{edsrcond}). The EDSR sub-harmonics on line (A) for $f=2.5$~GHz represent an example of isolated or ``pure'' EDSR and their spin dynamics will be considered in the next Sec. in time domain. It can be seen in Fig.\ref{figbzf} that at certain points in the $(B_z,f)$ plane the $\sigma_z^R$ maxima lines from all three families of lines merge with each other reflecting the condition $k_2=k_1+k_3$. We call the associated EDSR at such points as ``hybrid'' EDSR since the spin flip here is strongly mixed with tunneling into the neighboring dot. One can see that the points of ``hybrid'' EDSR are the hot spots in terms of the high average $\sigma_z^R$ amplitude so we may expect an effective spin flip in such regime. The spin evolution on the EDSR sub-harmonics for this case (line B) at $f=2.05$ GHz will also be studied in the next Sec. in time domain.

Panel (b) in Fig.~\ref{figbzf} shows the results from Eq.~(\ref{szrwa}) for the the time-averaged z-projection of the spin in the right QD obtained by the analytical approach within the RWA approximation described in Sec.~IIIB. One can see that the results of numerical (Fig.\ref{figbzf}a) and analytical (Fig.\ref{figbzf}b) approaches are in good agreement with each other on the EDSR sub-harmonics (\ref{edsrcond}) and even on the tunneling resonance lines (\ref{pat}) and (\ref{patflip}) which, strictly speaking, fall beyond the approximation (\ref{szrwa}). Such good agreement supports the role of dominant spin flip mechanisms introduced in our model.

\subsection{Spin flip map and EDSR sub-harmonics in the plane of magnetic field and tunneling rate}

\begin{figure}[tbp]
\centering
\includegraphics[width=0.48\textwidth]{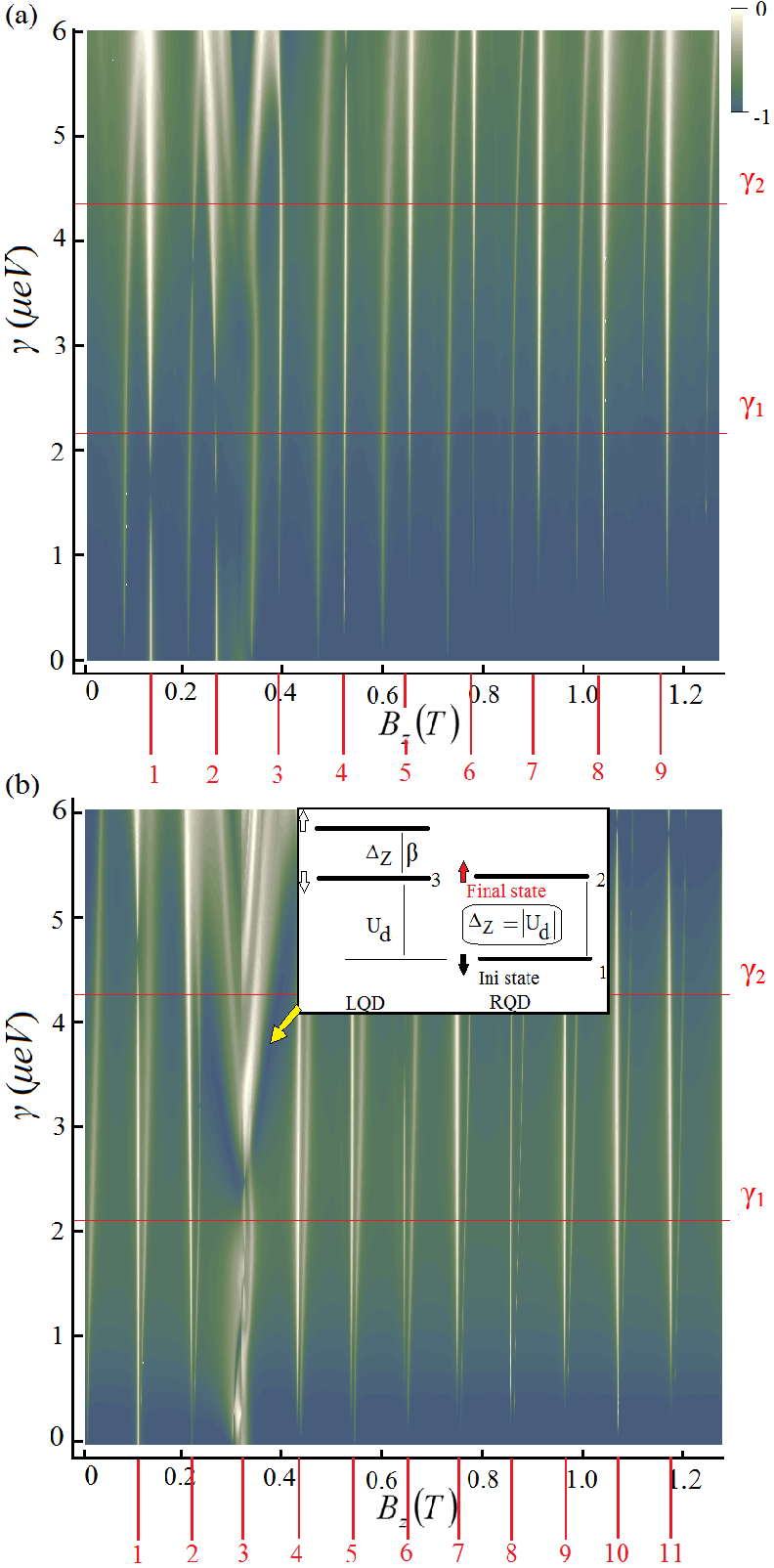}
\caption{Map for $\sigma_z^R$ in $(B_z, \gamma)$ coordinates for the driving frequency (a) $f=2.5$ GHz for line (A) in Fig.\ref{figbzf}(a) where the tunneling is not effective  at the condition corresponding to the EDSR sub-harmonics (numbered vertical lines) and (b) for $f=2.05$ GHz on line (B) in Fig.~\ref{figbzf}(a) where the EDSR is accompanied by effective tunneling between the dots due to PAT transitions). The other parameters are $U_d=-25 \mu eV$ and $V_d=75 \mu eV$. The horizontal red lines $(\gamma_1)$ and $(\gamma_2)$ are indicated positions for basic tunneling rate $\gamma_1=2.2$ $\mu eV$ and $\gamma_2=2 \gamma_1$ which are considered for the evolution examples in the text. The intensity of the EDSR sub-harmonics grows with increasing $\gamma$ which indicates the importance of tunneling for the spin evolution processes. The sidebands visible as curved lines near each EDSR sub-harmonic are the spin-flip tunneling lines corresponding to the condition (\ref{patflip}). The inset in panel (b) shows the level alignment for broadened sub-harmonic No.3 marked by yellow arrow when $\Delta_Z=|U_d|$ and the spin flip is strongly tunneling-dependent.}
\label{figbzgamma}
\end{figure}

To underline the role of tunneling in manifestation of the EDSR sub-harmonics, in Fig.\ref{figbzgamma} we show the spin flip map for $\sigma_z^R$ in the plane of $(B_z, \gamma)$ parameters representing the magnetic field and the tunneling rate. Panel (a) is for the driving frequency $f=2.5$ GHz 
corresponding to line (A) in Fig.\ref{figbzf}(a) where the EDSR condition (\ref{edsrcond}) is isolated from the PAT conditions (\ref{pat}) and (\ref{patflip}) so the tunneling is not effective although an increase in the tunneling rate speeds up the spin flip in the right QD. The EDSR sub-harmonics can be seen as vertical lines labeled by numbers below. The spin-flip PAT lines (\ref{patflip}) are also present between the EDSR lines as it can be seen also along line (A) in Fig.\ref{figbzf}(a). Panel (b) is for line (B) in Fig.\ref{figbzf}(a) at $f=2.05$ GHz where the EDSR is accompanied by effective tunneling since the conditions (\ref{pat}) - (\ref{edsrcond}) are satisfied together and all three lines merge together on EDSR sub-harmonics, as it can be seen along line (B) in Fig.\ref{figbzf}(a), creating the ``hybrid'' EDSR \cite{KS2022}. The inset in Fig.\ref{figbzgamma}b shows the level alignment for broadened sub-harmonic No.3 marked by yellow arrow when $\Delta_Z = |U_d|$ and the spin flip is strongly tunneling-dependent.
The other parameters are $U_d=-25~\mu eV$ and $V_d=30~\mu eV$. The horizontal red lines in Fig.~\ref{figbzgamma} are for ($\gamma_1$) basic tunneling rate $\gamma_1=2.18$ $\mu eV$ and ($\gamma_2$) $\gamma_2=2 \gamma_1$ which are considered for the evolution examples below in the text. The other fixed parameters are $U_d=-25~\mu eV$ and $V_d=30~\mu eV$. The tunneling rate $\gamma$ is varied between $0$ and $6$~$\mu eV$ corresponding to most of the structures used in the experiments \cite{Studenikin2018,Studenikin2021}. One can see that the intensity of the EDSR sub-harmonics (vertical lines) grows with increasing $\gamma$ which underlines the positive influence of tunneling to the neighboring on the spin flip at the dot where the spin has been initialized. The sidebands visible as curved lines near each of EDSR sub-harmonic are the spin-flip tunneling lines corresponding to the condition (\ref{patflip}). The distance between these sidebands and the vertical EDSR sub-harmonics grows with increasing tunneling rate $\gamma$ leading to a complicated anti-crossing-like behavior of the sideband curves at the top of the map.

\section{Evolution of the spin on EDSR sub-harmonics and spin flip efficiency}

Here we turn our attention to the stroboscopic evolution of the observables focusing on the spin projection (\ref{sigmazr}) in the right dot where the spin-down state has been initialized and where the main spin dynamics takes place. It is of practical interest to achieve controllable spin rotations both on main and higher resonance sub-harmonics which are considered in several papers \cite{StehlikPetta2014,Nowak2012,Szafran2017,Manchon2022}. Working with materials with large effective g-factor, large Zeeman splitting may not be accessible under the resonance condition (\ref{edsrcond}) on the main sub-harmonic $k_3=1$ due to the microwave hardware limitations but it might be accessible for some higher sub-harmonic $k_3 > 1$. Here we consider the spin evolution for the EDSR sub-harmonics visible on line (A) in Fig.\ref{figbzf} for ``pure'' EDSR where the tunneling is not effective and on line (B) for ``hybrid'' EDSR where the tunneling is very effective.

\subsection{Stroboscopic spin evolution on Bloch sphere}

The spin evolution in the EDSR regime on both main and higher sub-harmonics can be visualized by the stroboscopic evolution of the spin vector in the right QD \cite{KS2022},
\begin{equation}
{\bf S}^R(t_n)=(\sigma_x^R(t_n),\sigma_y^R(t_n),\sigma_z^R(t_n)),
\label{blochspin}
\end{equation}

\begin{figure*}[tbp]
\centering
\includegraphics*[width=0.67\textwidth]{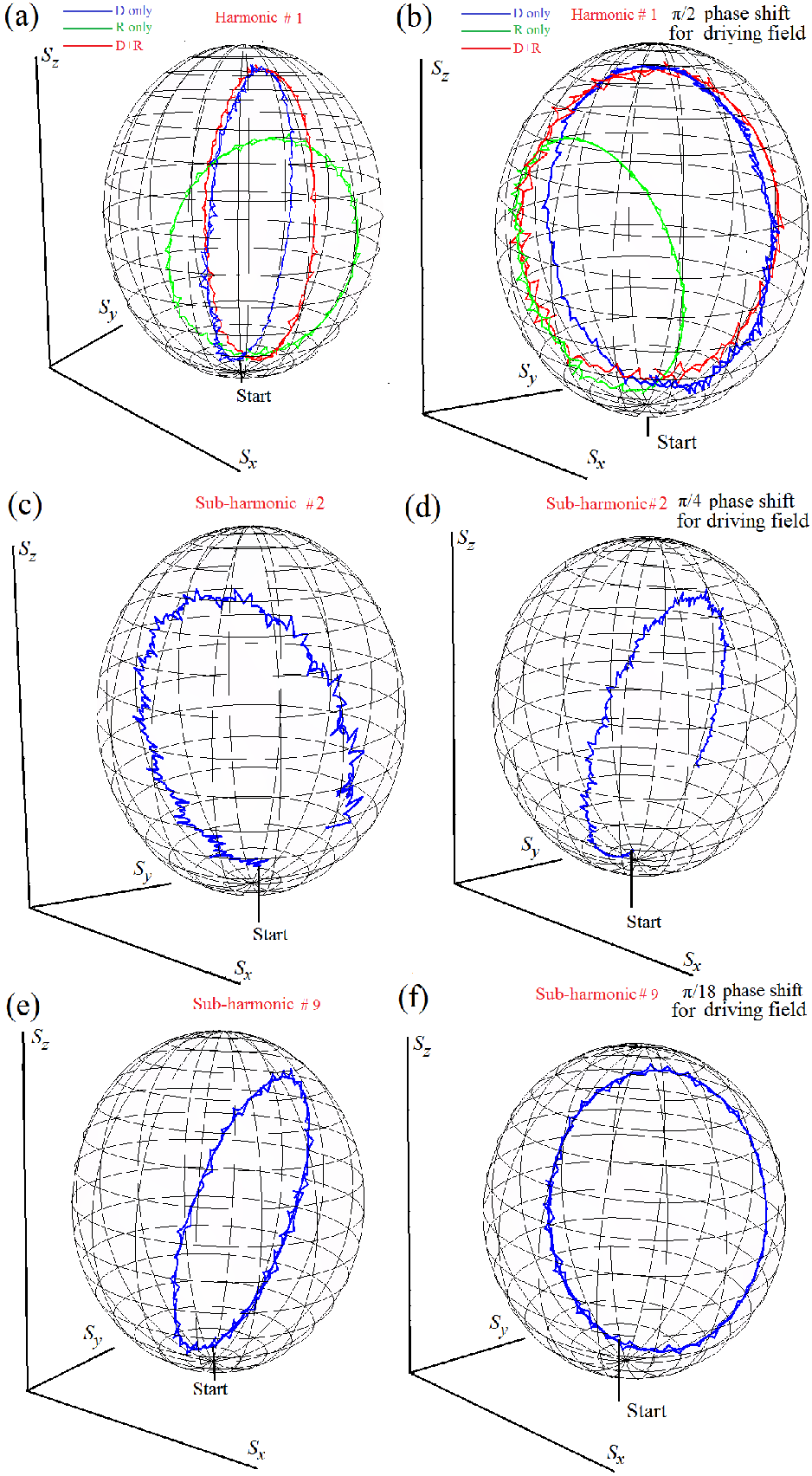}
\caption{Stroboscopic spin evolution for the right QD on the Bloch sphere shown for 200 driving periods corresponding to the EDSR main harmonic No.1 and its $k$-th sub-harmonic on line (A) in Fig.~\ref{figbzf}a for ``pure'' EDSR at $f=2.50$ GHz for (left) $\sin (\omega t)$ driving field and (right) $\sin (\omega t +\phi_k)$ driving field where the $\pi/2$ rotation of the spin evolution plane is achieved by the phase shift of the driving field $\phi_k=\pi/(2k)$: (a), (b) main harmonic $k=1$ where blue curve (D only) is for pure Dresselhaus SOC (\ref{hso}), green curve (R only) is for pure Rashba SOC (\ref{hr}) with the same amplitude, and red curve (D+R) is for combined presence of Dresselhaus and Rashba SOC; (c), (d) sub-harmonic $k=2$; (e), (f) sub-harmonic $k=9$, both plotted for pure Dresselhaus SOC. Arbitrary spin rotations around the x- and y-axes can be achieved for both main harmonic and the sub-harmonics.}
\label{figbloch1}
\end{figure*}

\begin{figure*}[tbp]
\centering
\includegraphics*[width=0.67\textwidth]{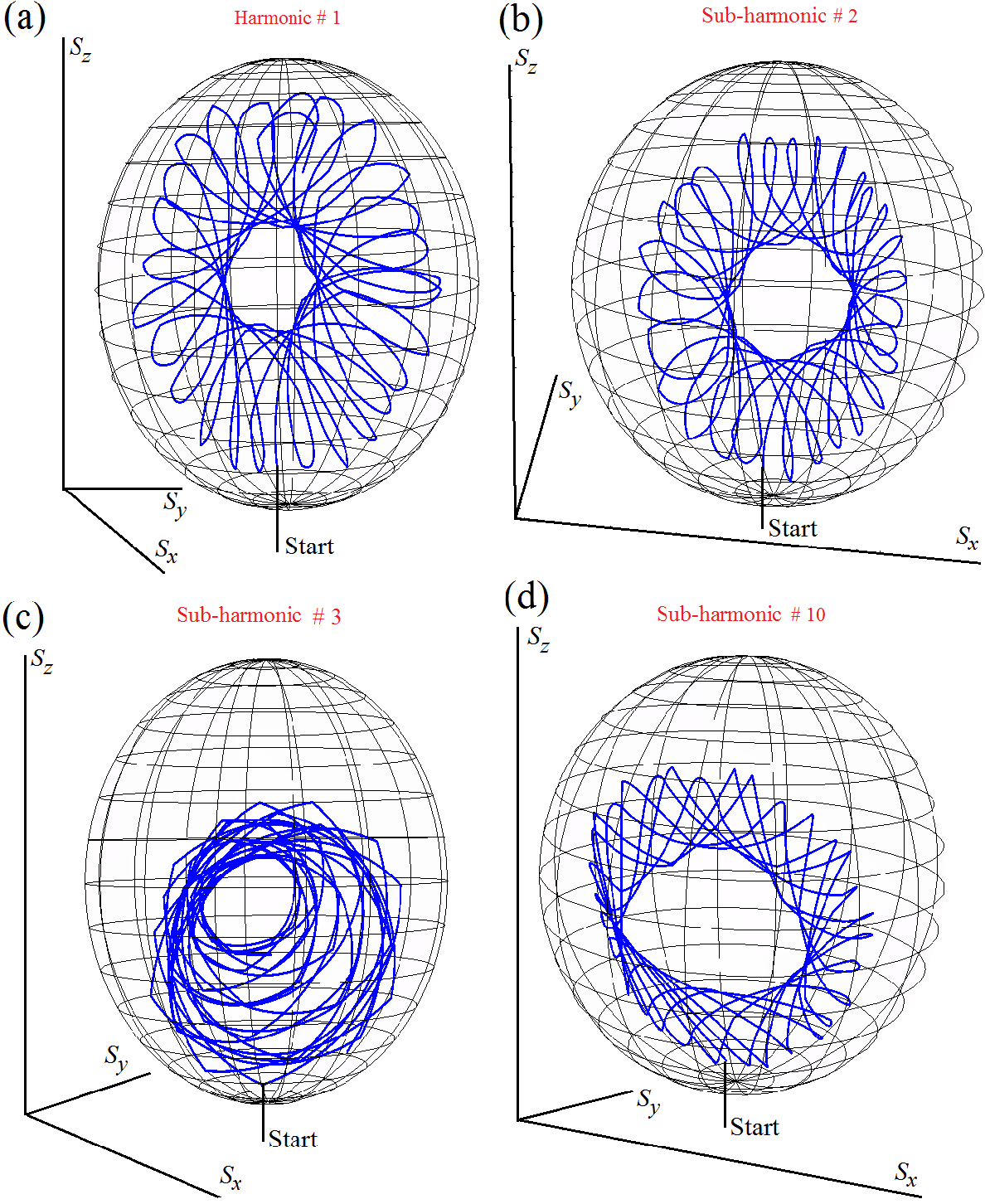}
\caption{Stroboscopic spin evolution on 200 driving periods for ``hybrid'' EDSR corresponding to the line (B) at $f=2.05$~GHz in Fig.~\ref{figbzf}a for (a) main harmonic $k=1$ and the sub-harmonics (b) $k=2$, (c) $k=3$ and (d) $k=10$. The tunneling is very effective and the Zeeman doublets in both QDs are equally involved into the dynamics, making the spin evolution patterns more complicated.}
\label{figbloch2}
\end{figure*}

where $t_n=nT$ is the stroboscopic time measured in the driving period $T$, the spin projection $\sigma_z^R(t)$ is calculated by (\ref{sigmazr}) and similar for the other projections in the right dot. For the qubit proposals it is important to have the controlled spin flip between the spin-down and spin-up states parallel to the direction of the magnetic field $B_z$ plus a possibility of $\pi/2$ rotations of the spin evolution plane around the $B_z$ axis. Such rotation for the $\sin (\omega t)$ driving field can be achieved at the $k$-th sub-harmonic by applying a specific phase shift to the field, $\sin (\omega t +\phi_k)$, where
\begin{equation}
\phi_k=\frac{\pi}{2k}.
\label{phaseshift}    
\end{equation}

The application of phase shift (\ref{phaseshift}) is necessary since the Larmour precession frequency is increased linearly with the sub-harmonic number $k$.

Naturally a question arises whether the Rashba contribution to SOC can produce any different or sizable effect on spin dynamics if one considers a nanostructure without the inversion symmetry of the quantum well potential. It is of interest to view the impact of pure Rashba SOC or its combination with the Dresselhaus term on the spin rotations on Bloch sphere. The structures with pure Rashba SOC can be found among Ge or Si quantum dots \cite{Petta2018,Vandersypen2015} where the inversion symmetry of the crystal structure prevents Dresselhaus term to arise, but the asymmetry of the macroscopic heterostructure potential leads to non-zero Rashba term. Below in Fig.\ref{figbloch1}(a),(b) we plot the  evolution of the spin vector (\ref{blochspin}) on the Bloch sphere on the main EDSR harmonic for three cases: pure Dresselhaus SOC (\ref{hso}) shown in blue color which is the basic case in our model; pure Rashba SOC 
\begin{equation}
H_R=\alpha_R k_x \sigma_y 
\label{hr}    
\end{equation}
instead of (\ref{hso}) with $\alpha_R=3$ $\rm{meV} \cdot \rm{nm}$ shown in green color where all the other parameters are the same as for the Dresselhaus SOC; combined Dresselhaus SOC (\ref{hso}) + Rashba SOC (\ref{hr}) shown by red color. 

It is a very challenging task to predict exact evolution of a four-level system encounting spin-orbit interactions, therefore, precise numerical simulations are helpful in  order to understand physics of the four-level system under the study.
The main difference between Dresselhaus and Rashba SOC in our effectively 1D model is that the Rashba term induces an effective magnetic field oriented approximately along the $Oy$ direction while the Dresselhaus term creates such field along the $Ox$ direction judging from the type of Pauli matrices. As a result, the spin rotation for pure Rashba term in ideal case would be expected to be around the $Oy$ axis while for pure Dresselhaus term it it would be around the $Ox$ axis. It should be mentioned, however, that in our double dot model  the situation is more complex and involves periodic tunneling events between the four levels. In particular, in qualitative terms the  hole moves periodically inside one dot or between the dots thus alternating the direction of the effective $k_x$-dependent magnetic field when $k_x$ is interchanged with $-k_x$. Besides, in a four-level system the dynamic patterns are defined by several SOC-induced matrix elements compared to a single element for a two-level system, making the evolution more complex and less predictable analytically. Another  complicating circumstance is that the evolution in Fig.\ref{figbloch1} and Fig.\ref{figbloch2} is shown in the stroboscopic fashion for the moments of time being multiples of the driving field period matching the Larmour rotation period around the $Oz$ axis which is parallel to the magnetic field. As a result, in continuous time we have the spin precession around at least two orthogonal directions oriented along the constant magnetic field and the $k_x$-dependent  effective SOC magnetic field. This makes the final effect of SOC on the spin vector trajectory (\ref{blochspin}) less obvious  compared to the basic case of the two-dimensional case with a simple SOC term. Namely, we may obtain the complex phase evolution for the spin rotations or even the modified radius of rotation on the Bloch sphere, especially for high sub-harmonics of the EDSR. This issue should be also addressed in the EDSR experiments where the precise match of the hardware stroboscopic window and the Larmour precession period can face various fluctuations which may lead to slow drift of the SOC-induced slow spin rotation plane.

The initial state is a spin-down state marked by the ``start'' label. We track the spin evolution for 200 driving periods on the main EDSR harmonic on panel (a) and we look on the effect of the $\pi/2$-phase shift of the driving field (\ref{phaseshift}) on panel (b). We see that the Rashba term alone with the same amplitude as the Dresselhaus term considered throughout the text makes similar $\pi$-rotation of the spin but around a different axis and with the smaller amplitude, spanning the smaller circle than the big circle of the sphere. The effect of the $\pi/2$-phase shift on panel (b) is the same as for the Dresselhaus term, i.e. the spin circling plane is rotated by $\pi/2$ around the direction of the magnetic field $B_z$. We conclude that the Rashba term alone makes a very similar effect on the spin rotation in our model but with somewhat weaker amplitude in the considered situation. Then, let us include both Rashba and Dresselhaus terms with equal amplitudes 
$\alpha_R=\beta_D=3$ $\rm{meV} \cdot \rm{nm}$. The results of the spin evolution for the main EDSR harmonic are shown in Fig.\ref{figbloch1}(a),(b) by the red curves. One can see that the effects on the spin trajectory of the combined presence of the Rashba and Dresselhaus terms are very similar to the effects of the sole Dresselhaus term presented by the blue curves including the $\pi/2$-rotation of the circling plane in panel (b). The reason is that the main effects of these SOC terms in 1D systems with localized states are in the nonzero matrix elements coupling the states with opposite spin by the scalar potential of the electrostatic field. Both, the Rashba and Dresselhaus terms produce similar effects for these matrix elements which actually enter the matrix evolution problem in the Hilbert space of the four localized states considered in our model. However, their time-dependent influence is modulated by the alternating direction of the hole movement inside one dot and between the dots and thus can be rather complex. We thus can conclude that the presence of the sole Dresselhaus term in the main part of our paper can be justified from the point of view that the addition of the Rashba term does not produce qualitatively new physics for the EDSR spin evolution on the Bloch sphere.

In Fig.\ref{figbloch1}(c)-(e) we show the stroboscopic evolution on 200 driving periods for the sub-harmonics $k=2$ and $k=9$, including the $\pi/2$-phase shift of the spin rotation plane on the right hand panels. 
It can be seen from Fig.\ref{figbloch1} that in the stroboscopic picture the down-up rotation of the spin is achieved both on main harmonic and on sub-harmonics within a single plane allowing the controlled spin operation for qubit proposals together with the $\pi/2$ rotation of the spin evolution plane. The stroboscopic nature of the evolution in the EDSR regime (\ref{edsrcond}) means that we move effectively to the rotating frame where the rotation frequency matches the Zeeman level splitting. It is known that the spin vector demonstrates an in-plane spin precession with such frequency. Such precession is not visible in the stroboscopic picture shown in Fig.\ref{figbloch1} but it can be useful for controlled in-plane spin rotations in possible qubit applications. 

The spin evolution for ``hybrid'' EDSR corresponding to the line (B) at $f=2.05$ GHz in Fig.\ref{figbzf}a is shown in Fig.\ref{figbloch2} for (a) main EDSR harmonic $k=1$ and for sub-harmonics (b) $k=2$, (c) $k=3$ and (d) $k=10$. Since in this regime the tunneling is very effective and the Zeeman doublets in both QDs are equally involved into the dynamics \cite{KS2022}, the spin evolution in Fig.\ref{figbloch2} obviously looks more complicated. Still, the large scale spin rotation can be achieved in this case also for the majority of the sub-harmonics.

\begin{figure*}[tbp]
\centering
\includegraphics*[width=0.9\textwidth]{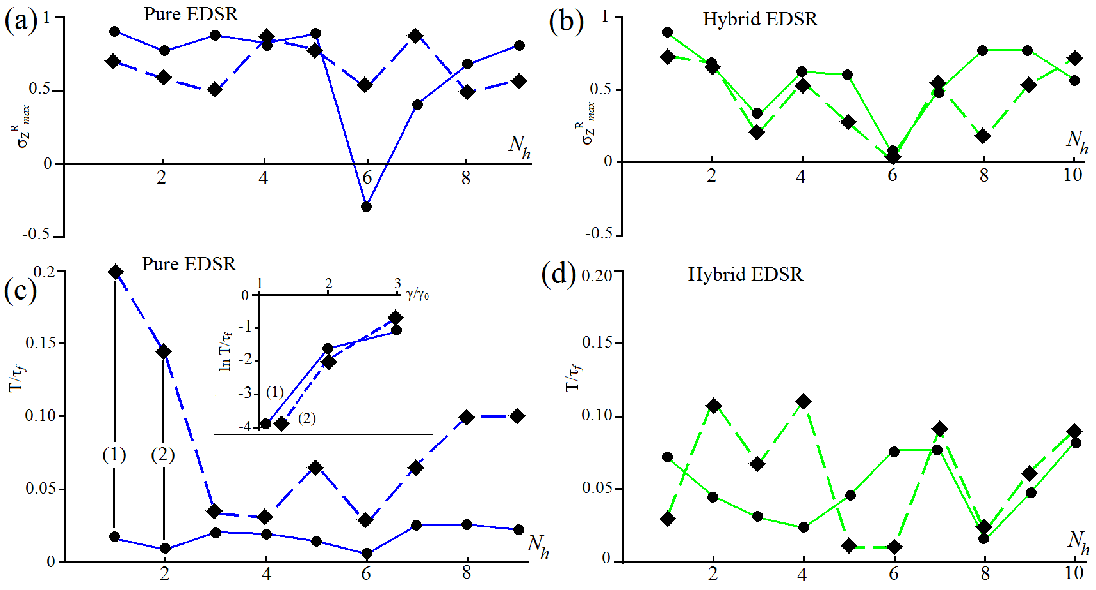}
\caption{(a), (b) Spin flip efficiency $\sigma_{Z {\rm max}}^R$ and (c), (d) inverse spin flip time $T/\tau_{f}$ vs EDSR sub-harmonic number $N_h$ extracted from the evolution data. Panels (a), (c) are for the "pure" EDSR with $f=2.5$ GHz corresponding to line A in Fig.\ref{figbzf} and panels (b), (d) are for the "hybrid" EDSR with $f=2.05$ GHz corresponding to line B in Fig.\ref{figbzf}. The circles connected by solid lines which serve as a guide to the eye are for basic tunnel coupling $\gamma_1=2.18$ $\mu eV$ and the diamonds connected by the dashed lines are for the enhanced coupling $\gamma=2 \gamma_1$. The spin flip amplitude is on average higher for the ``hybrid'' regime but the spin flip rate is enhanced most significantly with growing $\gamma$ for the first two sub-harmonics in ``pure'' regime labeled by vertical lines (1) and (2) in panel (c). For these sub-harmonics the inset in panel (c) shows the $\ln \tau_f/T$ vs the non-dimensional tunneling rate $x=\gamma/ \gamma_0$ with $\gamma_0=2.18$ $\mu eV$ for $x=1,2,3$ demonstrating the close to linear dependence.}
\label{figrates}
\end{figure*}

\subsection{Spin flip efficiency on EDSR sub-harmonics at different tunnel coupling}

It is of interest to find out how fast the spin flip can be achieved on both main and higher sub-harmonics of ``pure'' or ``hybrid'' EDSR  at different  values of the interdot tunnel coupling. We will focus on the associated characteristics of the spin evolution in this Subsection.
One may expect the significant effects of increased coupling on the spin flip efficiency which is of interest to explore on different EDSR sub-harmonics.
The spin evolution shown for EDSR harmonics of ``hybrid'' EDSR in Fig.\ref{figbloch2} demonstrates sometimes rather complicated shape and one should seek for an accessible and useful characteristic for it. It is important to know how deep the spin can be flipped from the initial spin-down state and how fast this can happen. To to this, we look at the maximum achieved $\sigma_{Z {\rm max}}^R$ directly from the raw evolution data and label the stroboscopic time $\tau_{f}$ at which $\sigma_{Z {\rm max}}^R$ is achieved. The ratio $T/\tau_{f}$ can serve as a spin flip rate measured in units of inverse driving period $1/T$. We register the values of  $\sigma_{Z {\rm max}}^R$ and $T/\tau_{f}$ for the majority of the sub-harmonics visible on lines (A) and (B) in Fig.\ref{figbzf}. The results for $\sigma_{Z {\rm max}}^R$ and $T/\tau_{f}$ are shown in Fig.\ref{figrates}. Panels (a), (b) show the spin flip efficiency $\sigma_{Z {\rm max}}^R$ vs sub-harmonic number $N_h$ and panels (c), (d) show the inverse spin flip time $T/\tau_{f}$ vs sub-harmonic number $N_h$, all being extracted from the raw evolution data. Panels (a), (c) are for the ``pure'' EDSR with $f=2.5$ GHz corresponding to line A in Fig.\ref{figbzf} and panels (b), (d) are for the ``hybrid'' EDSR with $f=2.05$ GHz corresponding to line B in Fig.\ref{figbzf}. The circles connected by solid lines which serve as a guide to the eye are for basic tunnel coupling $\gamma_1=2.18$ $\mu eV$ and the diamonds connected by dashed lines are for the enhanced coupling $\gamma=2 \gamma_1$.

By analyzing the results on Fig.\ref{figrates} one can see that the spin flip amplitude is on average higher for the "hybrid" regime, and the spin flip rate is enhance most significantly with growing $\gamma$ in the "pure" EDSR regime where its relative change is most significant. The spin flip inverse time $T/\tau_{sf}$ on panels (c), (d) demonstrates on average higher flip rates on higher sub-harmonics for the ``hybrid'' regime indicating a competitive advantage of the hybrid regime. Such advantage is, however, accompanied by a more complicated stroboscopic patterns of the spin dynamics shown in Fig.\ref{figbloch2}. Such complicated spin trajectories on/inside the Bloch sphere are typical for SOI-induced spin rotations especially when triggered by the driving pulses which have a shape different from a simple monochromatic driving \cite{Budagosky2016,Lasek2023}. 
 One may observe in Fig.\ref{figrates} that the spin flip efficiency and rate are both non-monotonous functions of the sub-harmonic number $N_h$. The spin flip can be comparably fast on basic and higher sub-harmonics which effective number manifested at the given driving frequency and strength for the specific Zeeman splitting is limited by conditions (\ref{maxk3}), (\ref{maxdeltaz}). However, at certain combinations of parameters one can fall into the ``dark'' area on a certain EDSR sub-harmonic line visible in Fig.\ref{figbzf} which can be illustrated by the oscillating character of the Bessel functions in analytical approximation (\ref{szrwa}) as a function of the driving strength $V_d$ and frequency $\omega$. Such example can be visible in Fig.\ref{figbzf}(a) for sub-harmonic No.6 when it crosses the lines at the selected frequencies $f=2.5$ GHz (line (A) in Fig.\ref{figbzf}(a)) and $f=2.05$ GHz (line (B) in Fig.\ref{figbzf}(a)). As a result, we observe a reduced spin flip efficiency and rate for sub-harmonic No.6 in Fig.\ref{figrates}.   One can conclude that the spin rotation efficiency can be optimized by moving along the EDSR sub-harmonics lines.
The inset in panel (c) shows the $\ln \tau_f/T$ vs the non-dimensional tunneling rate $x=\gamma/ \gamma_0$ with $\gamma_0=2.18$ $\mu eV$ for $x=1,2,3$ for the the first two sub-harmonics in panel (c) labeled by vertical lines (1) and (2) and demonstrating the most striking enhancement of the spin flip rate. In this inset we may observe the close to linear (in semi-logarithmic scale) dependence for $\ln \tau_f/T$. This is an indication of a strong, exponential-like dependence of the spin flip rate $\tau_f/T$ on the tunneling coupling which points to the tunneling enhancement as one of the effective  methods to control and optimise the spin manipulation rates.

\section{Conclusions}
 
We have studied both computationally and analytically the spin evolution of a four-level spin-resolved system formed in a double quantum dot in the presence of a strong spin-orbit interaction. We considered a spin qubit formed in the right quantum dot which behaviour is strongly affected by the tunneling coupling with the Zeeman-split levels in the neighboring (left) dot enhanced by the Landau-Zener-St{\"u}ckelberg-Majorana interference. We have shown that an arbitrary rotation of a spin-qubit formed in the right dot can be effectively performed at sub-harmonics of the electric dipole spin resonance around the x- and y- axes in the rotating Larmour precession frame.  
It was found that the spin flip rate in the primary dot is enhanced by tunnel coupling with the auxiliary dot at the main EDSR frequency and also at its high sub-harmonics allowing coherent spin $\pi$-rotations on a 10-ns time scale.
The proposals of spin manipulation on high sub-harmonics considered in our paper are promising for new time-efficient schemes of the spin control protocols in qubit devices with large Zeeman splitting, e.g., in high magnetic fields,  where the main harmonic frequency may be impractical due to hardware limitations.

\section*{Acknowledgements}

The authors are grateful to M.M. Glazov, S.A. Tarasenko, N.S. Averkiev, I.S. Burmistrov, L.E. Fedichkin, A.S. Melnikov, A.A. Andronov, I.D. Tokman for stimulating discussions. D.V.K., M.V.B. and N.A.Z. are supported by the Ministry of Science and Higher Education of the Russian Federation (State Assignment FSWR-2023-0035).

\end{document}